\documentclass[a4paper,11pt]{article}
\pdfoutput=1 % if your are submitting a pdflatex (i.e. if you have
\usepackage{jheppub} % for details on the use of the package, please
\usepackage{graphicx}
\usepackage{hyperref}
\usepackage{pdfpages}
\numberwithin{figure}{section}
\usepackage[section]{placeins}
\usepackage{epstopdf}

\usepackage[colorinlistoftodos]{todonotes}
\usepackage{epsfig}
\usepackage{graphicx,color}
\usepackage{cancel}
\usepackage[utf8]{inputenc}

\newcommand{\be}{\begin{equation}}
\newcommand{\ee}{\end{equation}}

\newcommand{\bea}{\begin{aligned}}
\newcommand{\eea}{\end{aligned}}

\newcommand{\bk}{{\bf k}}

\newcommand{\sL}{{\scriptscriptstyle L}}
\newcommand{\sT}{{\scriptscriptstyle T}}

\newcommand{\beq}{\begin{eqnarray}}
\newcommand{\eeq}{\end{eqnarray}}

\definecolor{red}{rgb}{1,0,0}
\definecolor{gray}{rgb}{0.5,0.5,0.5}

\renewcommand{\d}{\mathrm{d}}

\newcommand{\ef}{{\epsilon^2_f}}
\usepackage{bbold}
\usepackage{tensor}
\usepackage[normalem]{ulem} % \sout{old text} for strikeout
\renewcommand\sout{\bgroup  \ULdepth=-.5ex \ULset}

\newcommand{\bP}{\mathbf{P}}

\title{Low-$x$ improved TMD approach to the lepto- and hadroproduction of a heavy-quark pair}
\author[a]{Tolga Altinoluk,}
\author[b]{Cyrille Marquet,}
\author[b]{and Pieter Taels}

\affiliation[a]{National Centre for Nuclear Research, 02-093 Warsaw, Poland}
\affiliation[b]{Centre de Physique Th\'eorique, \'Ecole polytechnique, CNRS, I.P. Paris, F-91128 Palaiseau, France}

\emailAdd{tolga.altinoluk@ncbj.gov.pl}
\emailAdd{cyrille.marquet@polytechnique.edu}
\emailAdd{pieter.taels@polytechnique.edu}

\date{\today}

\preprint{???}

\abstract{

We study lepto- and hadroproduction of a heavy-quark pair in the ITMD factorization framework for dilute-dense collisions.
Due to the presence of a nonzero quark mass and/or nonzero photon virtuality, new contributions appear compared to the
cases of photo- and hadroproduction of dijets, for which the ITMD framework was originally derived.
The extra terms are sensitive to gluons that are not fully linearly polarized. At small $x$ those gluons emerge only when
saturation effects are taken into account, and in a proper way. As a result, in linear small-$x$ frameworks where gluon are fully linearly polarized,
such contributions are absent. We show however that they are not always negligible, even for large gluon transverse momentum, due
to the behavior of the off-shell hard factors.

}

\begin{document}

\maketitle

\section{Introduction}

Small-$x$ improved transverse-momentum-dependent (ITMD) factorization \cite{Kotko:2015ura,vanHameren:2016ftb} is the appropriate factorization scheme for describing small-$x$ dilute-dense processes that feature a hard transverse momentum $\mathbf{P}$ as well as a softer one $\mathbf{k}$. The prototypical example is forward dijet production proton-nucleus ($p+A$) collisions: the typical transverse momentum of a hard jet provides the hard scale, while the transverse momentum imbalance of the pair provides the other. In addition, the forward production imposes $x_p \gg x_{\scriptscriptstyle{A}}$, implying in turn a smaller intrinsic transverse momentum for the {\it dilute} projectile partons, of the order of $\Lambda_{QCD}$, compared to that of the {\it dense} target gluons, which is of the order of the target saturation momentum $Q_s(x_{\scriptscriptstyle{A}})$. Neglecting the former allows then to identify the latter with $\mathbf{k}$.

ITMD factorization is derived from the Color Glass Condensate (CGC) theory (see e.g. \cite{Gelis:2010nm}) in the $Q_s \ll |\mathbf{P}|$ limit, but keeping the $(Q_s/|\mathbf{k}|)^n$ and $(|\mathbf{k}|/|\mathbf{P}|)^n$ resummations intact \cite{Kotko:2015ura,Altinoluk:2019fui,Altinoluk:2019wyu,Boussarie:2020vzf}. As a result, ITMD expressions encompass other factorization approaches \cite{Dominguez:2010xd,Dominguez:2011wm,Iancu:2013dta,Kotko:2015ura,Marquet:2016cgx,Marquet:2017xwy}, namely $k_t$-factorization (also known as high-energy factorization (HEF)) \cite{Catani:1990eg,Deak:2009xt} and (the small-$x$ limit of) TMD factorization \cite{Angeles-Martinez:2015sea, Collins, Bomhof:2006dp}, that each have a narrower range of applicability associated to the value of $|\mathbf{k}|$: $Q_s \ll |\mathbf{k}|,|\mathbf{P}|$ and $Q_s,|\mathbf{k}| \ll |\mathbf{P}|$ respectively.

Schematically, an ITMD factorization formula reads
\be\bea
\mathrm{d\sigma} \propto f(x_p) \sum_{c} H^{ij}_{(c)}(\mathbf{P},\mathbf{k})
\Big[\frac{1}{2}\delta^{ij}\mathcal{F}_{\mathrm{(c)}}(x_{{\scriptscriptstyle A}},\bk)+\Big(\frac{k^{i}k^{j}}{\bk^{2}}-\frac{1}{2}\delta^{ij}\Big)\mathcal{H}_{\mathrm{(c)}}(x_{{\scriptscriptstyle A}},\bk)\Big]\;\label{eq:genericITMD}
\eea\ee
where $f(x_p)$ denotes a standard parton distribution function, $H^{ij}_{(c)}$ are hard factors with an off-shell small-$x_{\scriptscriptstyle{A}}$ gluon (but the large-$x_p$ parton is on-shell) and $\mathcal{F}_{\mathrm{(c)}}$ represents a set of process-dependent unpolarized transverse-momentum-dependent gluon distributions (gluon TMDs), with $\mathcal{H}_{\mathrm{(c)}}$ their linearly-polarized companions. Typically, given process will involve more than one $(\mathcal{F}_{\mathrm{(c)}},\mathcal{H}_{\mathrm{(c)}})$ pair of gluon TMDs.

All those distributions share a common perturbative tail known as the unintegrated gluon distribution (UGD), and the difference between them shows up at low values of $k_t$ where a single distribution ceases to be sufficient. Generically this is expected to happen in the non-perturbative regime. However at small values of $x_{\scriptscriptstyle{A}}$, $Q_s$ becomes large and takes over as the scale which characterises the soft rescatterings responsible for the duplication of the gluon distribution into several variants. Hence at small $x$, one may write $\mathcal{F}_{\mathrm{(c)}}, \mathcal{H}_{\mathrm{(c)}} = (1/\pi)\mathcal{F}_{g/A}+{\cal O}(Q^2_s/ \bk^{2})$ where $\mathcal{F}_{g/A}$ is the UGD; the difference between the gluon TMDs is therefore due to a resummation of non-linear corrections, called leading-twist saturation effects as they are not suppressed by powers of $\mathbf{P}^2$. The off-shell hard factors incorporate a resummation of so-called kinematical higher twists: $H^{ij}_{(c)} = H^{ij}_{(c)}(\mathbf{k}\!=\!0) + {\cal O}(\bk^{2}/\mathbf{P}^2) $ where $H^{ij}_{(c)}(\mathbf{k}\!=\!0)$ are the on-shell hard factors that appear at leading-twist. What is not included in the ITMD formula (but is in the full dilute-dense CGC expressions \cite{Marquet:2007vb,Dominguez:2011wm,Iancu:2013dta}) are the so-called genuine higher twists ${\cal O}(Q^2_s/\mathbf{P}^2)$ \cite{Altinoluk:2019wyu,Fujii:2020bkl}.

For completeness, let us add that neglecting kinematical twists by replacing the hard factors $H^{ij}_{(c)}$ with their collinear limit in \eqref{eq:genericITMD} yields TMD factorization (to be precise, TMD factorization in which only the small-$x$ gluon possesses nonzero transverse momentum). Conversely, neglecting saturation effects by replacing all the gluons TMDs by $\mathcal{F}_{g/A}/\pi$ yields high-energy factorization or $k_t$-factorization\footnote{note that using the HEF formula while obtaining $\mathcal{F}_{g/A}/\pi$ from the Balitsky-Kovchegov equation \cite{Balitsky:1995ub,Kovchegov:1999yj} instead of the Balitsky-Fadin-Kuraev-Lipatov \cite{Lipatov:1976zz,Kuraev:1976ge,Balitsky:1978ic} equation is not sufficient to correctly restore saturation effects; taking into account the various gluon TMDs is necessary as well.} (again, with only the small-$x$ gluon having nonzero transverse momentum).

One may alternatively write the ITMD formula in the following way
\be\bea
\frac{\mathrm{d\sigma}}{{\mathrm{d}^{3}\vec{k}_{1}\mathrm{d}^{3}\vec{k}_{2}}} \propto \sum_{c}
\Big[ H^{ns}_{(c)}(\mathbf{k}_1,\mathbf{k}_2) \mathcal{F}_{\mathrm{(c)}}(x_{{\scriptscriptstyle A}},\bk)+H^{h}_{(c)}(\mathbf{k}_1,\mathbf{k}_2) \Big(\mathcal{H}_{\mathrm{(c)}}(x_{{\scriptscriptstyle A}},\bk)-\mathcal{F}_{\mathrm{(c)}}(x_{{\scriptscriptstyle A}},\bk)\Big)\Big]\;
\eea\ee
with $H^{ns}_{(c)}=H^{ij}_{(c)}k^{i}k^{j}/\bk^{2}$ and $H^{h}_{(c)}=H^{ij}_{(c)}(k^{i}k^{j}/\bk^{2}-\delta^{ij}/2)$. The ITMD formula was first introduced for processes in which the hard factors $H^h_{(c)}$ vanish. That includes dijet production in $p+A$ collisions \cite{Kotko:2015ura} ($q g^*\to q g$, $g g^*\to q \bar{q}$, $g g^*\to gg$ processes) and photoproduction of dijets \cite{Kotko:2017oxg} ($\gamma g^*\to q \bar q$ process). However, in the presence of an additional scale -- such as a nonzero quark mass, a nonzero photon virtuality, or the presence of more production particles in the final state -- then $H^h_{(c)}\neq 0$ and as a result the fact that in general the unpolarized gluon TMDs $\mathcal{F}_{\mathrm{(c)}}$ differ from their linearly-polarized partners $\mathcal{H}_{\mathrm{(c)}}$ becomes relevant (a notable exception are gluon TMDs of the `dipole' type, fundamental ($\mathcal{F}_{\mathrm{DP}})$ or adjoint ($\mathcal{F}_{\mathrm{ADP}}$), for which $\mathcal{F}=\mathcal{H}$ \cite{Marquet:2017xwy})\footnote{as a consequence, a process like photon-jet production in $p+A$ collisions ($q g^*\to q \gamma^*$), which involves only $\mathcal{F}_{\mathrm{DP}}$, features no term $H^{h}_{(c)}(\mathcal{H}_{\mathrm{(c)}}\!-\!\mathcal{F}_{\mathrm{(c)}})$ and ITMD factorization in this case is nothing more than high-energy factorization.}.

The goal of the paper is to obtain the ITMD factorization formula for heavy-quark pair production in deep-inelastic scattering ($\gamma g^*\to Q \bar{Q}$ process) and $p+A$ collisions ($g g^*\to Q \bar{Q}$). In sections 2 and 3 we obtain the hard factors following the procedure introduced in \cite{Altinoluk:2019fui}, and in section 4 we compare the ITMD result to an approximation that we shall call ITMD*, in which one simply ignores the $H^{h}_{(c)}(\mathcal{H}_{\mathrm{(c)}}\!-\!\mathcal{F}_{\mathrm{(c)}})$ terms; such an approximation was used recently in the context of trijet production \cite{Bury:2020ndc}. In addition, for those cases where $H^h_{(c)}=0$, the hard factors that result from the so-called non-sense polarization projection $H^{ns}_{(c)}$ were first obtained from a diagrammatic approach involving only $2\to 2$ diagrams, in which the kinematical-twist resummation could be obtained by simply restoring the off-shellness of the incoming small-$x$ gluon. That approach however seems to be insufficient to obtain the $H^{h}_{(c)}$ projections for processes where they are nonzero. A second goal of the paper is to expose this apparent limitation of the diagrammatic approach, and this is the focus of section 5; we leave it for future work to study how could be altered to be made to work. Section 6 is devoted to conclusions and an outlook.

\section{\label{sec:DIS}Heavy-quark pair production in deep-inelastic scattering}
The differential cross section for the process $e(\ell)+A(p)\to e(\ell^{\prime})+Q(k_{1})+\bar{Q}(k_{2})+X$ in the correlation limit of the CGC is given by \cite{Metz:2011wb,Dominguez:2011br,Marquet:2017xwy}:
\be\bea
\frac{\mathrm{d}\sigma_{\ell A}}{\mathrm{d}x_{B}\mathrm{d}Q^{2}\mathrm{d}^{3}\vec{k}_{1}\mathrm{d}^{3}\vec{k}_{2}}  =\frac{\alpha e^2_Q}{\pi x_{B}Q^{2}}\Big[\frac{1+(1-y)^{2}}{2}\frac{\mathrm{d}\sigma_{\gamma_{T}^{*}A}}{\mathrm{d}^{3}\vec{k}_{1}\mathrm{d}^{3}\vec{k}_{2}}+(1-y)\frac{\mathrm{d}\sigma_{\gamma_{L}^{*}A}}{\mathrm{d}^{3}\vec{k}_{1}\mathrm{d}^{3}\vec{k}_{2}}\Big]\;,\label{eq:DIS}
\eea\ee
where the azimuthal angle of the outgoing lepton is integrated out\footnote{For this reason, there is no interference of the amplitudes induced by the transversely and longitudinally polarized virtual photon.} and where, for simplicity, we consider just one quark flavor.
In the above, we used the usual notation for the DIS invariants: $x_{B}=Q^{2}/2p\cdot q$, $y=p\cdot q/p\cdot\ell$, and $Q^2=-(\ell-\ell^\prime)^2$. The differential $\mathrm{d}^{3}\vec{k}_{1}\mathrm{d}^{3}\vec{k}_{2}$ in the three-momenta $\vec{k}_i\equiv (k^{+}_i,\mathbf{k}_i)$ of the outgoing quarks can be written as:
\be
\mathrm{d}^{3}\vec{k}_{1}\mathrm{d}^{3}\vec{k}_{2}=(q^+)^2\mathrm{d}z\mathrm{d}^{2}\mathbf{P}\,\mathrm{d}\bar{z}\mathrm{d}^{2}\mathbf{k}\;,
\ee
where $z=k_1^{+}/q^{+}$ and $\bar{z}=k_2^{+}/q^{+}$ are the longitudinal momentum fractions of the virtual photon carried by the (anti)quark. Moreover, we introduced the following combinations of transverse momenta:
\be \bea
\mathbf{P}\equiv\bar{z} \mathbf{k}_1 - z  \mathbf{k}_2\qquad \mathrm{and}\qquad \mathbf{k}\equiv\mathbf{k}_1+\mathbf{k}_2\;,
\eea\ee
used to extract the correlation limit (or TMD limit) by requiring $\mathbf{k}\ll \mathbf{P}$.

In that limit, the virtual photon-proton cross sections from eq. \eqref{eq:DIS} are found to be:
\be\bea
\left.\frac{\mathrm{d\sigma}_{\gamma_{\sT;\sL}^{*}A}}{\mathrm{d}^{3}\vec{k}_{1}\mathrm{d}^{3}\vec{k}_{2}}\right|_{TMD} & =\frac{\alpha \alpha_{s} e^2_Q}{(q^{+})^{2}}\delta(1-z-\bar{z})H^{ij}_{\sT;\sL}(\mathbf{P},\epsilon_f)\\
 & \times\Big[\frac{1}{2}\delta^{ij}\mathcal{F}_{\mathrm{WW}}(x_{{\scriptscriptstyle A}},\bk)+\frac{1}{2}\Big(2\frac{k^{i}k^{j}}{\bk^{2}}-\delta^{ij}\Big)\mathcal{H}_{\mathrm{WW}}(x_{{\scriptscriptstyle A}},\bk)\Big]\;,\label{eq:gammap}
\eea\ee
In the above cross sections, the hard parts $H^{ij}_{\sT;\sL}(\mathbf{P},\epsilon_f)$ depend on the hard momentum $\bP$ and on $\epsilon_f^2\equiv m^2+z \bar{z} Q^2$ with $m$ being the heavy-quark mass. Because of the colorless initial state photon, this process involves only one $(\mathcal{F}_{\mathrm{(c)}},\mathcal{H}_{\mathrm{(c)}})$ pair of gluon TMDs, which is of the Weizs\"acker-Williams (WW) type. All the dependence on the intrinsic momentum $\bk$ is absorbed into $\mathcal{F}_{\mathrm{WW}}$ and $\mathcal{H}_{\mathrm{WW}}$, and the hard parts that appear in eq. \eqref{eq:gammap} are given by\footnote{Throughout this work, we use the short-hand notations $\int_\mathbf{r}=\int\mathrm{d}^2\mathbf{r}$ and $\int_\mathbf{k}=\int\frac{\mathrm{d}^2\mathbf{k}}{(2\pi)^2}$ for integrals over transverse coordinates resp. momenta.}:
\be\bea
H^{ij}_{\sT}(\mathbf{P},\epsilon_f)&=\frac{1}{(2\pi)^2}\int_{\mathbf{r}\mathbf{r}^{\prime}}e^{-i\mathbf{P}\cdot(\mathbf{r}-\mathbf{r}^{\prime})}r^ir^{\prime j}\varphi_{s,s^\prime}^{\sT,\lambda}(\mathbf{r},\epsilon_{f})\varphi_{s,s^\prime}^{\sT,\lambda\dagger}(\mathbf{r}^{\prime},\epsilon_{f})\;,\label{eq:MT}
\eea\ee
and
\be\bea
H^{ij}_{\sL}(\mathbf{P},\epsilon_f)&=\frac{1}{(2\pi)^2}\int_{\mathbf{r}\mathbf{r}^{\prime}}e^{-i\mathbf{P}\cdot(\mathbf{r}-\mathbf{r}^{\prime})}r^ir^{\prime j}\varphi_{s,s^\prime}^{\sL}(\mathbf{r},\epsilon_{f})\varphi_{s,s^\prime}^{\sL\dagger}(\mathbf{r}^{\prime},\epsilon_{f})\;.\label{eq:ML}
\eea\ee
The products of the $\gamma_{\sT,\sL}^*\to Q \bar{Q}$ wave functions (summed over the quark spins $s,s^\prime$ and the photon's transverse polarization $\lambda$) are equal to:
\be\bea
\varphi_{s,s^\prime}^{\sT,\lambda}(\mathbf{r},\epsilon_{f})\varphi_{s,s^\prime}^{\sT,\lambda\dagger}(\mathbf{r}^{\prime},\epsilon_{f})&=(z^2+\bar{z}^2)\frac{\mathbf{r}\cdot\mathbf{r}^\prime}{|\mathbf{r}||\mathbf{r}^\prime|}\ef K_{1}(\epsilon_{f}|\mathbf{r}|)K_{1}(\epsilon_{f}|\mathbf{r}^\prime|)+m^2 K_{0}(\epsilon_{f}|\mathbf{r}|)K_{0}(\epsilon_{f}|\mathbf{r}^\prime|)\;,\\
\varphi_{s,s^\prime}^{\sL}(\mathbf{r},\epsilon_{f})\varphi_{s,s^\prime}^{\sL\dagger}(\mathbf{r}^{\prime},\epsilon_{f})&=4z^2\bar{z}^2Q^2 K_{0}(\epsilon_{f}|\mathbf{r}|)K_{0}(\epsilon_{f}|\mathbf{r}^\prime|)\;.\label{eq:overlap}
\eea\ee
Performing the integrations in eqs. \eqref{eq:MT} and \eqref{eq:ML}, one obtains the familiar results \cite{Boer:2016fqd}:
\be \bea
\left.\frac{\mathrm{d\sigma}_{\gamma_{\sT}^{*}}}{\mathrm{d}^{3}\vec{k}_{1}\mathrm{d}^{3}\vec{k}_{2}}\right|_{TMD} &=\frac{\alpha \alpha_{s} e^2_Q}{(q^{+})^{2}}\delta(1-z-\bar{z})\frac{1}{({\bf P}^{2}+\ef)^{4}}\\
&  \times\Big[\big((\bP^{4}+\epsilon_{f}^{4})(z^{2}+\bar{z}^{2})+2m^2 \bP^2\big)\mathcal{F}_{\mathrm{WW}}(x_{{\scriptscriptstyle A}},\bk)\\
&\qquad\qquad+\big(-2\bP^{2}\ef(z^{2}+\bar{z}^{2})+2m^2 \bP^2\big)\cos(2\varphi)\mathcal{H}_{\mathrm{WW}}(x_{{\scriptscriptstyle A}},\bk)\Big]\;,\label{eq:CGCTMDgammaT}
\eea\ee
where $\bP \cdot \bk=P\,k\,\mathrm{cos}(\varphi)$, and
\be \bea
\left.\frac{\mathrm{d\sigma}_{\gamma_{L}^{*}}}{\mathrm{d}^{3}\vec{k}_{1}\mathrm{d}^{3}\vec{k}_{2}}\right|_{TMD} & =\frac{\alpha \alpha_{s} e^2_Q}{(q^{+})^{2}}\delta(1-z-\bar{z})8z^2\bar{z}^2Q^2\frac{\bP^{2}}{(\bP^{2}+\ef)^{4}}\\
 & \times\Big[\mathcal{F}_{\mathrm{WW}}(x_{{\scriptscriptstyle A}},\bk)+\cos(2\varphi)\mathcal{H}_{\mathrm{WW}}(x_{{\scriptscriptstyle A}},\bk)\Big]\;.
\eea\ee
Explicit expressions for $\mathcal{F}_{\mathrm{WW}}$ and $\mathcal{H}_{\mathrm{WW}}$, (as well as for the other TMDs that we shall encounter below) are given in \cite{Marquet:2016cgx,Marquet:2017xwy}, along with their evaluation from simulations of the Jalilian-Marian-Iancu-McLerran-Weigert-Leonidov-Kovner (JIMWLK) \cite{JalilianMarian:1997jx,JalilianMarian:1997dw,Iancu:2000hn,Ferreiro:2001qy,Weigert:2000gi} equation.

\subsection{\label{sec:DISitmd}Reintroducing higher kinematical twists}
According to the analysis in refs. \cite{Altinoluk:2019fui,Boussarie:2020vzf}, the kinematical-twist corrections $(\bk^2/\bP^2)^{n}$ that are negligible in the TMD regime can be reintroduced and resummed in the hard parts by writing:
\be\bea
H^{ij}_{\sT}(z,\mathbf{k}_1,\mathbf{k}_2)&=\frac{1}{(2\pi)^2}\int_{\mathbf{r}\mathbf{r}^{\prime}}e^{-i\mathbf{P}\cdot(\mathbf{r}-\mathbf{r}^{\prime})}r^ir^{\prime j}\varphi_{s,s^\prime}^{\sT,\lambda}(\mathbf{r},\epsilon_{f})\varphi_{s,s^\prime}^{\sT,\lambda\dagger}(\mathbf{r}^{\prime},\epsilon_{f})\\
&\qquad \times \bigg( \frac{ e^{i\bar z {\bf k}\cdot {\bf r}}-e^{-iz {\bf k}\cdot {\bf r}}}{i  {\bf k\cdot {\bf r}}}\bigg)\bigg( \frac{ e^{-i\bar z {\bf k}\cdot {\bf r}^\prime}-e^{iz {\bf k}\cdot {\bf r}^\prime}}{-i {\bf k\cdot {\bf r}^\prime}}\bigg)\;,\label{eq:MTitmd}
\eea\ee
and
\be\bea
H^{ij}_{\sL}(z,\mathbf{k}_1,\mathbf{k}_2)&=\frac{1}{(2\pi)^2}\int_{\mathbf{r}\mathbf{r}^{\prime}}e^{-i\mathbf{P}\cdot(\mathbf{r}-\mathbf{r}^{\prime})}r^ir^{\prime j}\varphi_{s,s^\prime}^{\sL}(\mathbf{r},\epsilon_{f})\varphi_{s,s^\prime}^{\sL\dagger}(\mathbf{r}^{\prime},\epsilon_{f})\\
&\qquad \times \bigg( \frac{ e^{i\bar z {\bf k}\cdot {\bf r}}-e^{-iz {\bf k}\cdot {\bf r}}}{i  {\bf k\cdot {\bf r}}}\bigg)\bigg( \frac{ e^{-i\bar z {\bf k}\cdot {\bf r}^\prime}-e^{iz {\bf k}\cdot {\bf r}^\prime}}{-i {\bf k\cdot {\bf r}^\prime}}\bigg)\;.\label{eq:MLitmd}
\eea\ee
To evaluate these hard parts explicitly, we first cast them in the following form:
\be\bea
H^{ij}_{\sT}(z,\mathbf{k}_1,\mathbf{k}_2)&=(z^2+\bar{z}^2) \mathcal{M}^{i\lambda}_{1}(z,\mathbf{k}_1,\mathbf{k}_2)\mathcal{M}^{j\lambda\dagger}_{1}(z,\mathbf{k}_1,\mathbf{k}_2)+ m^2 \mathcal{M}^{i}_{0}(z,\mathbf{k}_1,\mathbf{k}_2)\mathcal{M}^{j\dagger}_{0}(z,\mathbf{k}_1,\mathbf{k}_2)\;,\\
H^{ij}_{\sL}(z,\mathbf{k}_1,\mathbf{k}_2)&=4z^2\bar{z}^2Q^2\mathcal{M}^{i}_{0}(z,\mathbf{k}_1,\mathbf{k}_2)\mathcal{M}^{j\dagger}_{0}(z,\mathbf{k}_1,\mathbf{k}_2)\;,
\label{eq:decomphard}\eea\ee
with the amplitudes:
\be\bea
\mathcal{M}^{i}_{0}(z,\mathbf{k}_1,\mathbf{k}_2)=\frac{1}{2\pi}\int_{\mathbf{r}}e^{-i\mathbf{P}\cdot\mathbf{r}}r^i K_0(\epsilon_{f}|\mathbf{r}|)\bigg( \frac{ e^{i\bar z {\bf k}\cdot {\bf r}}-e^{-iz {\bf k}\cdot {\bf r}}}{i  {\bf k\cdot {\bf r}}}\bigg)\;,\label{eq:M0}
\eea\ee
and
\be\bea
\mathcal{M}^{i\lambda}_{1}(z,\mathbf{k}_1,\mathbf{k}_2)=\frac{\epsilon_f}{2\pi}\int_{\mathbf{r}}e^{-i\mathbf{P}\cdot\mathbf{r}}\frac{r^i r^\lambda}{|\mathbf{r}|} K_{1}(\epsilon_{f}|\mathbf{r}|)\bigg( \frac{ e^{i\bar z {\bf k}\cdot {\bf r}}-e^{-iz {\bf k}\cdot {\bf r}}}{i  {\bf k\cdot {\bf r}}}\bigg)\;.\label{eq:M1}
\eea\ee

Let us start with the amplitude eq. \eqref{eq:M1}. Writing the Bessel function in momentum space, we obtain:
\beq
\frac{\epsilon_f}{2\pi}\frac{r^i r^\lambda}{|\mathbf{r}|} K_{1}(\epsilon_{f}|\mathbf{r}|)=  \int_\mathbf{L} \: e^{i{\bf L}\cdot {\bf r}}
\bigg(\frac{\delta^{i\lambda}}{{\bf L}^2+\ef}-2 \frac{{\bf L}^i{\bf L}^\lambda}{({\bf L}^2+\ef)^2}\bigg)\;.
\eeq
Combining the above with the trick:
\beq
\label{trick_1}
\frac{e^{i{\bf k}\cdot {\bf r}}-1}{i{\bf k}\cdot{\bf r}}=\int_0^1 \d t \; e^{it {\bf k}\cdot{\bf r}}\;,
\eeq
the amplitude can be cast in the following form:
\beq
\mathcal{M}^{i\lambda}_{1}=\int_0^1 \d t \; \bigg(\frac{\delta^{i\lambda}}{({\bf P}+z{\bf k}-t{\bf k})^2+\ef}-2\frac{({\bf P}+z{\bf k}-t{\bf k})^i({\bf P}+z{\bf k}-t{\bf k})^\lambda}{[({\bf P}+z{\bf k}-t{\bf k})^2+\ef]^2}\bigg)\; .
\eeq
After a little bit of algebra, one arrives at the following expression for the amplitude squared:
\be\bea
 \mathcal{M}_{1}^{i\lambda} \mathcal{M}_{1}^{j\lambda\dagger} &=\frac{1}{{\bf k}^{4}}I_{1}^{2}\delta^{ij}-\frac{4\ef}{{\bf k}^{8}}\Big({\bf P}^{i}I_{4}-{\bf k}^{i}I_{T}\Big)\Big({\bf P}^{j}I_{4}-{\bf k}^{j}I_{T}\Big)\\
	&+\frac{1}{\mathbf{P}^{2}\sin^{2}\varphi+\ef}\Big(\frac{1}{(\mathbf{k}_{1}^{2}+\ef)(\mathbf{k}_{2}^{2}+\ef)}-\frac{I_{1}^{2}}{{\bf k}^{4}}\Big)\\
	&\qquad\times\bigg[{\bf P}^{i}{\bf P}^{j}+(\mathbf{P}^{2}+\ef)\frac{k^{i}k^{j}}{\mathbf{k}^{2}}-\frac{{\bf k}\cdot{\bf P}}{\mathbf{k}^{2}}({\bf k}^{i}{\bf P}^{j}+{\bf P}^{i}{\bf k}^{j})\bigg]\;.
\eea\ee
The integrals that appear are the following:
\be\bea
I_1&=\frac{|\mathbf{k}|}{\sqrt{\mathbf{P}^{2}\sin^{2}\varphi+\ef}}\bigg[\arctan\bigg(\frac{{\bf k}\cdot\mathbf{k}_{1}}{|\mathbf{k}|\sqrt{\mathbf{P}^{2}\sin^{2}\varphi+\ef}}\bigg)\\
&\qquad\qquad\qquad\qquad+\arctan\bigg(\frac{{\bf k}\cdot\mathbf{k}_{2}}{|\mathbf{k}|\sqrt{\mathbf{P}^{2}\sin^{2}\varphi+\ef}}\bigg)\bigg]\;,
\eea\ee
\be\bea
I_4&=\frac{1}{2}\frac{\mathbf{k}^{2}}{\mathbf{P}^{2}\sin^{2}\varphi+\ef}\bigg[\frac{{\bf k}\cdot\mathbf{k}_{1}}{\mathbf{k}_{1}^{2}+\ef}+\frac{{\bf k}\cdot\mathbf{k}_{2}}{\mathbf{k}_{2}^{2}+\ef}+I_{1}\bigg]\;
\eea\ee
and
\be\bea
I_{\sT}
&=\frac{1}{2}\frac{\mathbf{k}^{2}}{\mathbf{P}^{2}\sin^{2}\varphi+\ef}\bigg[\frac{-z{\bf k}\cdot\mathbf{k}_{1}}{\mathbf{k}_{1}^{2}+\ef}+\frac{\bar{z}{\bf k}\cdot\mathbf{k}_{2}}{\mathbf{k}_{2}^{2}+\ef}+\frac{{\bf k}\cdot{\bf P}}{\mathbf{k}^{2}}I_{1}\bigg]\;.
\eea\ee
Finally, projecting on the tensor structures that multiply the gluon TMDs in eq. \eqref{eq:gammap}, we obtain:
\be\bea
\mathcal{M}_{1}^{i\lambda} \mathcal{M}_{1}^{j\lambda\dagger} \frac{k^{i}k^{j}}{{\bf k}^{2}}&=\frac{1}{\mathbf{P}^{2}\sin^{2}\varphi+\epsilon_{f}^{2}}\Bigg[\frac{\mathbf{P}^{2}\sin^{2}\varphi}{(\mathbf{k}_{1}^{2}+\epsilon_{f}^{2})(\mathbf{k}_{2}^{2}+\epsilon_{f}^{2})}+\frac{\epsilon_{f}^{2}}{{\bf k}^{4}}\Big(\frac{{\bf k}\cdot\mathbf{k}_{1}}{\mathbf{k}_{1}^{2}+\epsilon_{f}^{2}}+\frac{{\bf k}\cdot\mathbf{k}_{2}}{\mathbf{k}_{2}^{2}+\epsilon_{f}^{2}}\Big)^{2}\Bigg]\;,\label{eq:HTns}
\eea\ee
and
\be\bea
\mathcal{M}_{1}^{i\lambda}\mathcal{M}_{1}^{j\lambda\dagger}\frac{\delta^{ij}}{2}&=\frac{1}{(\mathbf{k}_{1}^{2}+\epsilon_{f}^{2})(\mathbf{k}_{2}^{2}+\epsilon_{f}^{2})}\frac{\mathbf{P}^{2}\sin^{2}\varphi}{\mathbf{P}^{2}\sin^{2}\varphi+\epsilon_{f}^{2}}\\
	&-\frac{\epsilon_{f}^{2}}{\mathbf{P}^{2}\sin^{2}\varphi+\epsilon_{f}^{2}}\frac{I_{1}}{{\bf k}^{4}}\bigg(\frac{{\bf k}\cdot\mathbf{k}_{1}}{\mathbf{k}_{1}^{2}+\epsilon_{f}^{2}}+\frac{{\bf k}\cdot\mathbf{k}_{2}}{\mathbf{k}_{2}^{2}+\epsilon_{f}^{2}}\bigg)\\
	&+\frac{\epsilon_{f}^{4}}{2{\bf k}^{4}}\frac{1}{(\mathbf{P}^{2}\sin^{2}\varphi+\epsilon_{f}^{2})^{2}}\bigg(\frac{{\bf k}\cdot\mathbf{k}_{1}}{\mathbf{k}_{1}^{2}+\epsilon_{f}^{2}}+\frac{{\bf k}\cdot\mathbf{k}_{2}}{\mathbf{k}_{2}^{2}+\epsilon_{f}^{2}}+I_{1}\bigg)^{2}\;.\label{eq:HTf}
\eea\ee

The amplitude $\mathcal{M}_{0}$ in eq. \eqref{eq:M0}, can be treated in a similar way. One obtains:
\be\bea
\mathcal{M}_{0}^{i}=\frac{1}{\mathbf{P}^{2}\sin^{2}\varphi+\ef}\frac{1}{\mathbf{k}^{2}}\bigg[\frac{{\bf k}\cdot\mathbf{k}_{1}}{\mathbf{k}_{1}^{2}+\ef}{\bf k}_{1}^{i}-\frac{{\bf k}\cdot\mathbf{k}_{2}}{\mathbf{k}_{2}^{2}+\ef}{\bf k}_{2}^{i}+I_{1}{\bf P}^{k}\bigg(\delta^{ki}-\frac{{\bf k}^{k}{\bf k}^{i}}{{\bf k}^{2}}\bigg)\bigg]\;.
\eea\ee
Squaring the above result and projecting on $k^{i}k^{j}/{\bf k}^{2}$ yields: 
\be\bea
\mathcal{M}_{0}^{i}\mathcal{M}_{0}^{\dagger j}\frac{k^{i}k^{j}}{{\bf k}^{2}}&= \frac{1}{{\bf k}^{2}}\Big(\frac{\mathbf{k}^2_{1}-\mathbf{k}^2_{2}}{(\mathbf{k}_{1}^{2}+\ef)(\mathbf{k}_{2}^{2}+\ef)}\Big)^{2}\;.\label{eq:HLns}
\eea\ee
Likewise, the projection of the amplitude squared on the tensor structure $\delta^{ij}/2$:
\be\bea
\mathcal{M}_{0}^{i}\mathcal{M}_{0}^{\dagger j}\frac{\delta^{ij}}{2}&=
\frac{1}{2{\bf k}^{4}}\frac{1}{\mathbf{P}^{2}\sin^{2}\varphi+\ef}\bigg[\frac{{\bf k}^{4}}{(\mathbf{k}_{1}^{2}+\ef)(\mathbf{k}_{2}^{2}+\ef)}+2I_{1}\Big(\frac{{\bf k}\cdot\mathbf{k}_{1}}{\mathbf{k}_{1}^{2}+\ef}+\frac{{\bf k}\cdot\mathbf{k}_{2}}{\mathbf{k}_{2}^{2}+\ef}\Big)\\
&\qquad+I_{1}^{2}-\frac{\ef}{\mathbf{P}^{2}\sin^{2}\varphi+\ef}\Big(\frac{{\bf k}\cdot\mathbf{k}_{1}}{\mathbf{k}_{1}^{2}+\ef}+\frac{{\bf k}\cdot\mathbf{k}_{2}}{\mathbf{k}_{2}^{2}+\ef}+I_{1}\Big)^{2}\bigg]\;.\label{eq:HLf}
\eea\ee

Collecting the above results, we can write down our final result for the $\gamma^*A\to Q\bar{Q}X$ cross sections in the ITMD framework:
\be\bea
\frac{\mathrm{d\sigma}_{\gamma_{\sT,\sL}^{*}}}{\mathrm{d}z\mathrm{d}^{2}\mathbf{k}_{1}\mathrm{d}^{2}\mathbf{k}_{2}} & =\alpha \alpha_{s} e^2_Q\Big[H^f_{\sT,\sL}(z,\mathbf{k}_1,\mathbf{k}_2) \mathcal{F}_{\mathrm{WW}}(x_{{\scriptscriptstyle A}},\bk)+H^h_{\sT,\sL}(z,\mathbf{k}_1,\mathbf{k}_2) \mathcal{H}_{\mathrm{WW}}(x_{{\scriptscriptstyle A}},\bk)\Big]\\
& =\alpha \alpha_{s} e^2_Q\Big[H^{ns}_{\sT,\sL}(z,\mathbf{k}_1,\mathbf{k}_2) \mathcal{F}_{\mathrm{WW}}(x_{{\scriptscriptstyle A}},\bk)\\
&\hspace{2cm} +H^h_{\sT,\sL}(z,\mathbf{k}_1,\mathbf{k}_2)
\Big(\mathcal{H}_{\mathrm{WW}}(x_{{\scriptscriptstyle A}},\bk)-\mathcal{F}_{\mathrm{WW}}(x_{{\scriptscriptstyle A}},\bk)\Big)\Big]\;,\label{eq:DISITMD}
\eea\ee
with:
\be\bea
H^f_{\sT,\sL}(z,\mathbf{k}_1,\mathbf{k}_2) &\equiv H^{ij}_{\sT,\sL}(z,\mathbf{k}_1,\mathbf{k}_2)\frac{\delta^{ij}}{2}\;,\\
H^h_{\sT,\sL}(z,\mathbf{k}_1,\mathbf{k}_2) &\equiv H^{ij}_{\sT,\sL}(z,\mathbf{k}_1,\mathbf{k}_2)\Big(\frac{k^{i}k^{j}}{{\bf k}^{2}}-\frac{\delta^{ij}}{2}\Big)\;,\\
H^{ns}_{\sT,\sL}(z,\mathbf{k}_1,\mathbf{k}_2) &= H^f_{\sT,\sL}(z,\mathbf{k}_1,\mathbf{k}_2)+H^h_{\sT,\sL}(z,\mathbf{k}_1,\mathbf{k}_2)\;.\label{eq:DISITMDhard}
\eea\ee

We checked that in the HEF limit $\mathbf{k}^2\gg Q^2_s$ where all TMDs coincide with the UGD, the expression
\be\bea
\left.\frac{\mathrm{d\sigma}_{\gamma_{\sT,\sL}^{*}}}{\mathrm{d}z\mathrm{d}^{2}\mathbf{k}_{1}\mathrm{d}^{2}\mathbf{k}_{2}}\right|_{HEF} &=\frac{\alpha \alpha_{s} e^2_Q}{\pi}
H^{ns}_{\sT,\sL}(z,\mathbf{k}_1,\mathbf{k}_2)\mathcal{F}_{g/A}(x_{{\scriptscriptstyle A}},\bk)\label{eq:HEFdis}
\eea\ee
coincides with the HEF result in \cite{Catani:1990eg} (appendix B).

In the limit where $m\to0$ and $Q^2\to 0$, we have that $H^{h}_{\sT} \to 0$, as expected. In that case, the only difference between the ITMD expression for the $\gamma A\to q \bar{q} X$ cross-section $\mathrm{d\sigma}_{\gamma}/\mathrm{d}z\mathrm{d}^{2}\mathbf{k}_{1}\mathrm{d}^{2}\mathbf{k}_{2}\!=\!\alpha \alpha_{s} e^2_Q (z^2\!+\!\bar{z}^2) \mathcal{F}_{\mathrm{WW}}(x_{{\scriptscriptstyle A}},\bk) /(\mathbf{k}_1^2\mathbf{k}_2^2) $ and the HEF one, is the presence of $\mathcal{F}_{\mathrm{WW}}$ in the former and of $\mathcal{F}_{g/A}/\pi$ in the latter. This is because for this particular process, only one type of gluon TMDs is involved. However, when extending the HEF formula to include saturation effects, one should keep in mind that $\mathcal{F}_{g/A}/\pi$ should indeed be replaced by $\mathcal{F}_{\mathrm{WW}}$, and not by $\mathcal{F}_{\mathrm{DP}}$ (the fundamental dipole TMD). This is crucial because those two distributions behave very differently at low values of $|\bk|$.

\section{\label{sec:pA}Forward heavy-quark pair production in proton-nucleus scattering}

The calculation of the kinematical resummation for the $gA\to Q\bar{Q}X$ is performed along the same lines as the previous case, however in that case there is more than one $(\mathcal{F}_{\mathrm{(c)}},\mathcal{H}_{\mathrm{(c)}})$  pair of gluon TMDs involved. Specifically, three different types play a role: the Weizs\"acker-Williams (WW), the adjoint dipole (ADP), and third type denoted type (1) \cite{Marquet:2017xwy}. Because the unpolarized and linearly polarized adjoint dipole gluon TMDs are identical, an $H_{\mathrm{ADP}}^{h}$ hard factor does not appear in the formula, and furthermore the WW- and (1)-type TMDs share the same hard parts. The ITMD cross section therefore may be written:
\be\bea
\frac{\mathrm{d}\sigma_{gA}}{\mathrm{d}z\mathrm{d}^{2}\mathbf{k}_{1}\mathrm{d}^{2}\mathbf{k}_{2}} & =\frac{\alpha_{s}^{2}}{4C_{F}}
\Big[H_{\mathrm{ADP}}^{ns}\mathcal{F}_{\mathrm{ADP}}(x_{{\scriptscriptstyle A}},\mathbf{k})+
H_{1}^{ns}\Big(\mathcal{F}_{gg}^{(1)}(x_{{\scriptscriptstyle A}},\mathbf{k})-\frac{1}{N_{c}^{2}}\mathcal{F}_{\mathrm{WW}}(x_{{\scriptscriptstyle A}},\mathbf{k})\Big)\\
 & +H_{1}^{h}\Big(\mathcal{H}_{gg}^{(1)}(x_{{\scriptscriptstyle A}},\mathbf{k})-\mathcal{F}_{gg}^{(1)}(x_{{\scriptscriptstyle A}},\mathbf{k})-\frac{1}{N_{c}^{2}}\mathcal{H}_{\mathrm{WW}}(x_{{\scriptscriptstyle A}},\mathbf{k})+\frac{1}{N_{c}^{2}}\mathcal{F}_{\mathrm{WW}}(x_{{\scriptscriptstyle A}},\mathbf{k})\Big)\Big]\;.\label{eq:sigmagA}
\eea\ee

The hard factors are obtained from:
\be\bea
H^{ij}_{1}(z,\mathbf{k}_1,\mathbf{k}_2)&=\frac{1}{(2\pi)^2}\int_{\mathbf{r}\mathbf{r}^{\prime}}e^{-i\mathbf{P}\cdot(\mathbf{r}-\mathbf{r}^{\prime})}r^ir^{\prime j}\
\varphi_{s,s^\prime}^{\sT,\lambda}(\mathbf{r},m)\varphi_{s,s^\prime}^{\sT,\lambda\dagger}(\mathbf{r}^{\prime},m)\\
&\qquad \times \bigg( \frac{ e^{i\bar z {\bf k}\cdot {\bf r}}-e^{-iz {\bf k}\cdot {\bf r}}}{i  {\bf k\cdot {\bf r}}}\bigg)\bigg( \frac{ e^{-i\bar z {\bf k}\cdot {\bf r}^\prime}-e^{iz {\bf k}\cdot {\bf r}^\prime}}{-i {\bf k\cdot {\bf r}^\prime}}\bigg)
\eea\ee
and
\be\bea
H_{\mathrm{ADP}}^{ij}(z,\mathbf{k}_1,\mathbf{k}_2)&=\frac{1}{(2\pi)^2}\int_{\mathbf{r}\mathbf{r}^{\prime}}e^{-i\mathbf{P}\cdot(\mathbf{r}-\mathbf{r}^{\prime})}r^ir^{\prime j}\
\varphi_{s,s^\prime}^{\sT,\lambda}(\mathbf{r},m)\varphi_{s,s^\prime}^{\sT,\lambda\dagger}(\mathbf{r}^{\prime},m) \\
&\qquad \times 
\Big[\bigg( \frac{e^{i\bar z {\bf k}\cdot {\bf r}}-1}{i  {\bf k\cdot {\bf r}}}\bigg)
\bigg( \frac{e^{iz {\bf k}\cdot {\bf r}^\prime}-1}{-i {\bf k\cdot {\bf r}^\prime}}\bigg)
+
\bigg( \frac{e^{-i z {\bf k}\cdot {\bf r}}-1}{i  {\bf k\cdot {\bf r}}}\bigg)
\bigg( \frac{e^{-i \bar z {\bf k}\cdot {\bf r}^\prime}-1}{-i {\bf k\cdot {\bf r}^\prime}}\bigg)\Big]
\eea\ee
where the $g\to Q \bar{Q}$ wave function overlap is given by \eqref{eq:overlap} but with $\epsilon_{f}$ replaced by $m$ (as we've factorized the couplings and charges out).

Projecting onto the usual 2-d Lorentz tensor we then obtain $H_{1}^{ns}=H_{1}^{ij} k^{i}k^{j}/\mathbf{k}^{2}$ and $H_{1}^{h}=H_{1}^{ij}(k^{i}k^{j}/\mathbf{k}^{2}-\delta^{ij}/2)$:
\be\bea
H_{1}^{ns}&=\frac{z^2+\bar{z}^2}{(\mathbf{k}_{1}^{2}+m^{2})(\mathbf{k}_{2}^{2}+m^{2})}\\
&+\frac{2z\bar{z}m^{2}}{\mathbf{k}^{6}\big(\mathbf{P}^{2}\sin^{2}\varphi+m^{2}\big)^{2}}\Big(\frac{({\bf k}\cdot\mathbf{k}_{1})^{2}}{\mathbf{k}_{1}^{2}+m^{2}}-\frac{({\bf k}\cdot{\bf k}_{2})^{2}}{\mathbf{k}_{2}^{2}+m^{2}}\Big)^{2}\;.\label{eq:H1ns}
\eea\ee
and
\begin{align}
H_{1}^{h}&=\frac{z^2+\bar{z}^2}{2(\mathbf{k}_{1}^{2}+m^{2})(\mathbf{k}_{2}^{2}+m^{2})}\Big(1-\frac{\mathbf{P}^{2}\sin^{2}\varphi}{\mathbf{P}^{2}\sin^{2}\varphi+m^{2}}\Big)\nonumber\\
&-\frac{m^{2}(z^2+\bar{z}^2)}{2\mathbf{k}^{4}\big(\mathbf{P}^{2}\sin^{2}\varphi+m^{2}\big)}I_{1}^{2}+\frac{2z\bar{z}m^{2}}{\mathbf{k}^{6}\big(\mathbf{P}^{2}\sin^{2}\varphi+m^{2}\big)^{2}}\Big(\frac{({\bf k}\cdot\mathbf{k}_{1})^{2}}{\mathbf{k}_{1}^{2}+m^{2}}-\frac{({\bf k}\cdot{\bf k}_{2})^{2}}{\mathbf{k}_{2}^{2}+m^{2}}\Big)^{2}\nonumber\\
&-\frac{z\bar{z}m^{2}}{\mathbf{k}^{4}\big(\mathbf{P}^{2}\sin^{2}\varphi+m^{2}\big)^{2}}\Big(\frac{{\bf k}\cdot\mathbf{k}_{1}}{\mathbf{k}_{1}^{2}+m^{2}}\mathbf{k}_{1}-\frac{{\bf k}\cdot\mathbf{k}_{2}}{\mathbf{k}_{2}^{2}+m^{2}}\mathbf{k}_{2}+\big(\mathbf{P}-\frac{{\bf k}\cdot\mathbf{P}}{{\bf k}^{2}}\mathbf{k}\big)I_{1}\Big)^{2}\;,\label{eq:H1h}
\end{align}

%\begin{align}
%H_{1}^{ij}\frac{\delta^{ij}}{2}&=\frac{P_{qg}(z)}{(\mathbf{k}_{1}^{2}+m^{2})(\mathbf{k}_{2}^{2}+m^{2})}\Big(1+\frac{\mathbf{P}^{2}\sin^{2}\varphi}{\mathbf{P}^{2}\sin^{2}\varphi+m^{2}}\Big)%\nonumber\\
%&+\frac{m^{2}P_{qg}(z)}{\mathbf{k}^{4}\big(\mathbf{P}^{2}\sin^{2}\varphi+m^{2}\big)}I_{1}^{2}\\
%&+\frac{z\bar{z}m^{2}}{\mathbf{k}^{4}\big(\mathbf{P}^{2}\sin^{2}\varphi+m^{2}\big)^{2}}\Big(\frac{{\bf k}\cdot\mathbf{k}_{1}}{\mathbf{k}_{1}^{2}+m^{2}}\mathbf{k}_{1}-\frac{{\bf k}\cdot\mathbf{k}_{2}}{\mathbf{k}_{2}^{2}+m^{2}}\mathbf{k}_{2}+\big(\mathbf{P}-\frac{{\bf k}\cdot\mathbf{P}}{{\bf k}^{2}}\mathbf{k}\big)I_{1}\Big)^{2}\;,\nonumber
%\end{align}

The hard part corresponding to the ADP TMD read:
\be\bea
H_{\mathrm{ADP}}^{ns}&=H_{\mathrm{ADP}}^{ij}\frac{k^{i}k^{j}}{\mathbf{k}^{2}} =\frac{2\bar{z}z(z^2+\bar{z}^2)}{\mathbf{P}^{2}+m^{2}}\frac{\mathbf{k}_{1}\cdot\mathbf{k}_{2}-m^{2}}{(\mathbf{k}_{1}^{2}+m^{2})(\mathbf{k}_{2}^{2}+m^{2})}\\
 & +\frac{4z\bar{z}m^{2}}{\mathbf{k}^{6}\big(\mathbf{P}^{2}\sin^{2}\varphi+m^{2}\big)^{2}}\Big(\frac{({\bf k}\cdot\mathbf{P})^{2}}{\mathbf{P}^{2}+m^{2}}-\frac{({\bf k}\cdot\mathbf{k}_{1})^{2}}{\mathbf{k}_{1}^{2}+m^{2}}\Big)\Big(\frac{({\bf k}\cdot\mathbf{P})^{2}}{\mathbf{P}^{2}+m^{2}}-\frac{({\bf k}\cdot{\bf k}_{2})^{2}}{\mathbf{k}_{2}^{2}+m^{2}}\Big)\;.
 \label{eq:Hadp}
 \eea\ee
while the other is not needed ($\mathcal{F}_{\mathrm{ADP}}=\mathcal{H}_{\mathrm{ADP}}$). As before, in the massless limit $m\to 0$, $H_1^{h}$ vanishes (and for completeness so does $H_{\mathrm{ADP}}^{h}$), and we find agreement with the ITMD result in ref.~\cite{Kotko:2015ura}.
%\be\bea
%H_{\mathrm{ADP}}^{ij}\frac{\delta^{ij}}{2}= & \frac{2P_{qg}(z)z\bar{z}}{\mathbf{P}^{2}+m^{2}}\frac{\mathbf{k}_{1}\cdot\mathbf{k}_{2}-m^{2}}{(\mathbf{k}_{1}^{2}+m^{2})(\mathbf{k}_{2}^{2}+m^{2})}\Big(1+\frac{\mathbf{P}^{2}\sin^{2}\varphi}{\mathbf{P}^{2}\sin^{2}\varphi+m^{2}}\Big)\\
 %& -\frac{m^{2}}{\mathbf{k}^{4}\big(\mathbf{P}^{2}\sin^{2}\varphi+m^{2}\big)}2P_{qg}(z)I_{1}^{(z)}I_{1}^{(\bar{z})}\\
 %& +\frac{2\bar{z}zm^{2}}{\mathbf{k}^{4}\big(\mathbf{P}^{2}\sin^{2}\varphi+m^{2}\big)^{2}}\Big(\frac{{\bf k}\cdot\mathbf{P}}{\mathbf{P}^{2}+m^{2}}\mathbf{P}-\frac{{\bf k}\cdot\mathbf{k}_{1}}{\mathbf{k}_{1}^{2}+m^{2}}\mathbf{k}_{1}-\big(\mathbf{P}-\frac{{\bf k}\cdot\mathbf{P}}{\mathbf{P}^{2}}\mathbf{k}\big)I_{1}^{(z)}\Big)\\
 %& \quad\times\Big(\frac{{\bf k}\cdot\mathbf{P}}{\mathbf{P}^{2}+m^{2}}\mathbf{P}-\frac{{\bf k}\cdot\mathbf{k}_{2}}{\mathbf{k}_{2}^{2}+m^{2}}\mathbf{k}_{2}+\big(\mathbf{P}-\frac{{\bf k}\cdot\mathbf{P}}{\mathbf{P}^{2}}\mathbf{k}\big)I_{1}^{(\bar{z})}\Big)\;,
%\eea\ee

As required, the cross section eq.~\eqref{eq:sigmagA} simplifies into the TMD-factorized result in the limit $\mathbf{k} \to 0$: 
\begin{align}
\left.\frac{\mathrm{d}\sigma_{gA}}{\mathrm{d}z\mathrm{d}^{2}\mathbf{k}_{1}\mathrm{d}^{2}\mathbf{k}_{2}}\right|_{TMD}  =\frac{\alpha_{s}^{2}}{2C_{F}}
&\frac{1}{\big(\mathbf{P}^{2}+m^{2}\big)^{2}}
\Bigg\{\Big(\frac{z^2+\bar{z}^2}2+\frac{2z\bar{z}m^{2}\mathbf{P}^{2}}{\big(\mathbf{P}^{2}+m^{2}\big)^{2}}\Big)\nonumber\\
&\times\Big(\mathcal{F}_{gg}^{(1)}(x_{{\scriptscriptstyle A}},\mathbf{k})-2z\bar{z}\mathcal{F}_{\mathrm{ADP}}(x_{{\scriptscriptstyle A}},\mathbf{k})-\frac{1}{N_{c}^{2}}\mathcal{F}_{\mathrm{WW}}(x_{{\scriptscriptstyle A}},\mathbf{k})\Big)\nonumber\\
 & +\frac{2z\bar{z}m^{2}\mathbf{P}^{2}}{\big(\mathbf{P}^{2}+m^{2}\big)^{2}}\cos(2\varphi)\\
 &\times\Big(\mathcal{H}_{gg}^{(1)}(x_{{\scriptscriptstyle A}},\mathbf{k})-2z\bar{z}\mathcal{F}_{\mathrm{ADP}}(x_{{\scriptscriptstyle A}},\mathbf{k})-\frac{1}{N_{c}^{2}}\mathcal{H}_{\mathrm{WW}}(x_{{\scriptscriptstyle A}},\mathbf{k})\Big)\Bigg\}\;\nonumber,
\end{align}
in agreement with refs.~\cite{Marquet:2017xwy,Akcakaya:2012si}.

Conversely, in the HEF limit $\mathbf{k}^2\gg Q^2_s$, where all TMDs, unpolarized or linearly polarized, coincide with the unintegrated gluon distribution (UGD):
\be\bea
\mathcal{F}_{gg}^{(1)}(x_{{\scriptscriptstyle A}},\mathbf{k}),\, \mathcal{F}_{\mathrm{ADP}}(x_{{\scriptscriptstyle A}},\mathbf{k}),\,\mathcal{F}_{\mathrm{WW}}(x_{{\scriptscriptstyle A}},\mathbf{k}),\,\mathcal{H}_{gg}^{(1)}(x_{{\scriptscriptstyle A}},\mathbf{k}),\, \mathcal{H}_{\mathrm{WW}}(x_{{\scriptscriptstyle A}},\mathbf{k})\to\,\frac{1}{\pi}\mathcal{F}_{g/A}(x_{{\scriptscriptstyle A}},\mathbf{k})\;,
\eea\ee
the ITMD cross section eq.~\eqref{eq:sigmagA} then becomes
\be\bea
\left.\frac{\mathrm{d}\sigma_{gA}}{\mathrm{d}z\mathrm{d}^{2}\mathbf{k}_{1}\mathrm{d}^{2}\mathbf{k}_{2}}\right|_{HEF} & =\frac{\alpha_{s}^{2}}{4\pi C_{F}}\Big((1-1/N_c^2)H_{1}^{ns}+H_{\mathrm{ADP}}^{ns}\Big)\mathcal{F}_{g/A}(x_{{\scriptscriptstyle A}},\mathbf{k})\;,\label{eq:pAHEF}
\eea\ee
in agreement with the HEF expressions \cite{Catani:1990eg} (appendix B).

\section{Numerical comparison between ITMD* and ITMD frameworks}

In \cite{farid}, the $\gamma^* A \to q\bar q X$ ITMD expression \eqref{eq:DISITMD} is compared to the full CGC expression, similarly to what was done in \cite{Fujii:2020bkl} for the $gA \to q\bar q X$ case, in order to study the genuine twists contributions, which are parametrically suppressed by powers of $Q^2_s/\mathbf{P}^2$ and neglected in the ITMD framework, but included in the CGC.

We instead concentrate on a different comparison: we would like to compare our formulae to an approximation which appeared recently in the literature, dubbed ITMD*, which consists in neglecting the $H^{h}_{(c)}(\mathcal{H}_{\mathrm{(c)}}\!-\!\mathcal{F}_{\mathrm{(c)}})$ terms in the ITMD expressions. That approximation had to be invoked in the context of trijet production \cite{Bury:2020ndc}, for which the $H^{h}_{(c)}$ hard factors are still unknown. The many unpolarized gluons TMDs involved in that process were obtained in \cite{Bury:2018kvg}, and the associated $H^{ns}_{(c)}$ matrix elements were computed numerically in \cite{Bury:2020ndc}. From studies of the TMD limit of other three-particle processes \cite{Altinoluk:2018uax,Altinoluk:2018byz,Altinoluk:2020qet}, it is known that the missing $H^{h}_{(c)}$'s are nonzero, and we would like to assess how important they are, in processes where we do know them.

Because the difference between the ITMD* and ITMD frameworks involves $\mathcal{H}_{\mathrm{(c)}}\!-\!\mathcal{F}_{\mathrm{(c)}}$, one would expect the former to be a good approximation of the latter when $|\mathbf{k}|$ becomes large, since all TMDs share the same perturbative limit, however we will see that this is not necessarily the case.
Using the gluon TMDs obtained in \cite{Marquet:2017xwy} from numerical simulations of the JIMWLK equations, we present the ratios of the cross sections for heavy-quark pair lepto- and hadroproduction calculated in the ITMD* approximation and in the ITMD framework.

In all plots, we choose\footnote{we purposely avoid choosing $|\mathbf{k}_1|=|\mathbf{k}_2|$ to stay away from accidental simplications, e.g. $H^{ns}_{\sL}=0$.} $|\mathbf{k}_1|=\sqrt{10}\,\mathrm{GeV}$, $|\mathbf{k}_2|=\sqrt{11}\,\mathrm{GeV}$, $z=\bar{z}=1/2$, and $m=m_c=1.275\,\mathrm{GeV}$. For leptoproduction, we choose a virtuality of $Q^2=10\,\mathrm{GeV}^2$. The ratios are presented as a function of $\alpha$, the angle between the transverse momenta $\bk_1$ and $\bk_2$ of the outgoing massive quarks, implying that $|\mathbf{k}|$ is small near $\alpha=\pi$ and grows as one goes away from that back-to-back configuration. We also use three different values of $x_{\scriptscriptstyle{A}}$, corresponding to (gluon) saturation scales of $Q_s(x_{\scriptscriptstyle{A}}=10^{-2})\approx0.6$, $Q_s(10^{-3})\approx0.95$, and  $Q_s(10^{-4})\approx1.5$ $\mathrm{GeV}$ (we are using $\alpha_s=0.2$ to translate the evolution steps into $\ln(1/x_{\scriptscriptstyle{A}})$ increments).

\begin{figure}[t] 
\centering
\includegraphics[width=11cm]{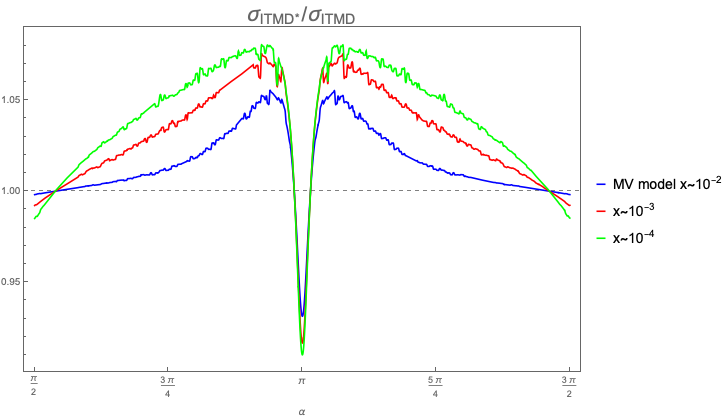} 
\caption{Ratio of the $\gamma^{*}_{\sT}A\to Q\bar{Q}X$ cross sections calculated in the ITMD* and ITMD frameworks, as a function of the angle $\alpha$ between the transverse momenta of the heavy quarks. Gluon TMDs are numerically evaluated in the McLerran-Venugopalan (MV) model \cite{MV}, and evolved towards lower values of $x$ with the JIMWLK equations.}
 \label{fig:DIST}
\end{figure}

\begin{figure}[t] 
\centering
\includegraphics[width=11cm]{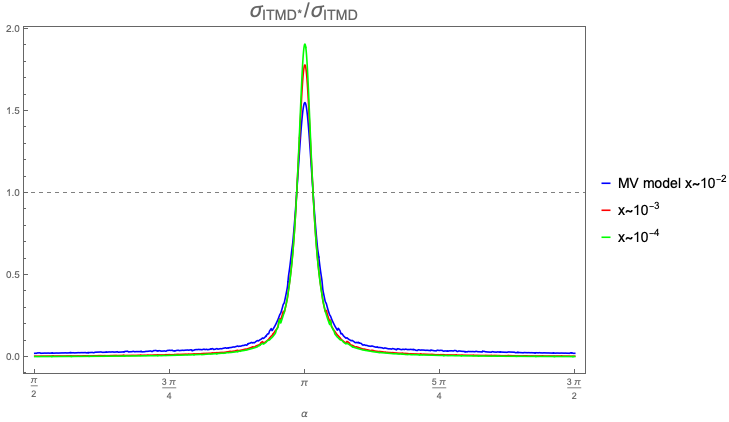} 
\caption{Ratio of the $\gamma^{*}_{\sL}A\to Q\bar{Q}X$ cross sections calculated in the ITMD* and ITMD frameworks, as a function of the angle $\alpha$ between the transverse momenta of the heavy quarks. Gluon TMDs are numerically evaluated in the MV model, and evolved towards lower values of $x$ with the JIMWLK equations.}
 \label{fig:DISL}
\end{figure}

\begin{figure}[t] 
\centering
\includegraphics[width=11cm]{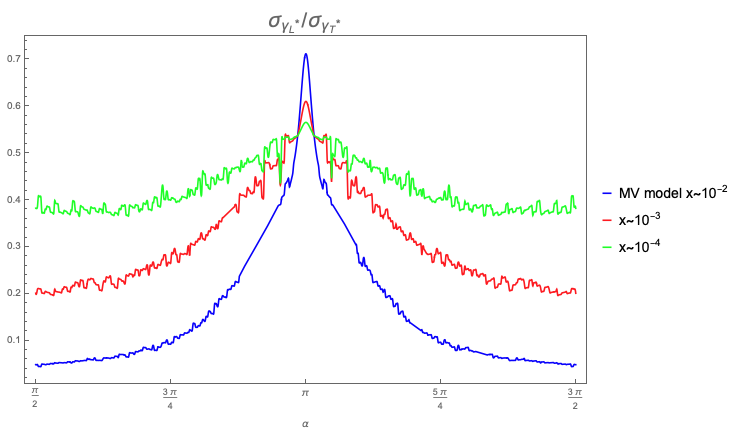} 
\caption{Ratio of the $\gamma^{*}_{\sL}A\to Q\bar{Q}X$ and $\gamma^{*}_{\sT}A\to Q\bar{Q}X$ cross sections, calculated in the ITMD framework and as a function of the angle $\alpha$ between the transverse momenta of the heavy quarks. Gluon TMDs are numerically evaluated in the MV model, and evolved towards lower values of $x$ with the JIMWLK equations.}
 \label{fig:ratioLT}
\end{figure}

In figs.~\ref{fig:DIST} and \ref{fig:DISL}, we present the ratios of the $\gamma^{*}_{\sT,\sL}A\to Q\bar{Q}X$ cross sections calculated as (suppressing, for ease of notation, the dependencies in the hard parts and TMDs):
\be\bea
\frac{\mathrm{d\sigma}_{\gamma_{\sT,\sL}^{*}}(\mathrm{ITMD^*})}{\mathrm{d\sigma}_{\gamma_{\sT,\sL}^{*}}(\mathrm{ITMD})} & =\frac{H^{ns}_{\sT,\sL} \mathcal{F}_{\mathrm{WW}}}{H^{ns}_{\sT,\sL}\mathcal{F}_{\mathrm{WW}}+H^{h}_{\sT,\sL}\big( \mathcal{H}_{\mathrm{WW}}-\mathcal{F}_{\mathrm{WW}}\big)}\;.
\eea\ee
For the process initiated by a transversely polarized virtual photon, fig.~\ref{fig:DIST}, the difference between both schemes is initially minor, around 5\%, and slightly increases in importance when evolving towards lower $x$. However, in the case of $\gamma^{*}_{\sL}A\to Q\bar{Q}X$, presented in fig.~\ref{fig:DISL}, the difference between ITMD* and ITMD is dramatic. Away from the back-to-back region, as a result of the kinematical-twist resummation, the hard part ratio $H^{ns}_{\sL}/H^{h}_{\sL}$ goes to zero faster than the TMD ratio $(\mathcal{H}_{\mathrm{WW}}-\mathcal{F}_{\mathrm{WW}})/\mathcal{F}_{\mathrm{WW}}$, making the $H^{h}_{\sL}$ term dominant, and ITMD* a very bad approximation. In the back-to-back configuration, the ITMD result is overestimated by a factor of 50 to 100\%, depending on the evolution of the gluon TMDs. Except for photoproduction, when $y=1$, this $\gamma^{*}_\sL$-contribution to the leptoproduction cross section is not negligible compared to $\gamma^{*}_\sT$, as illustrated in fig.~\ref{fig:ratioLT}.

\begin{figure}[t]
\centering
\includegraphics[width=11cm]{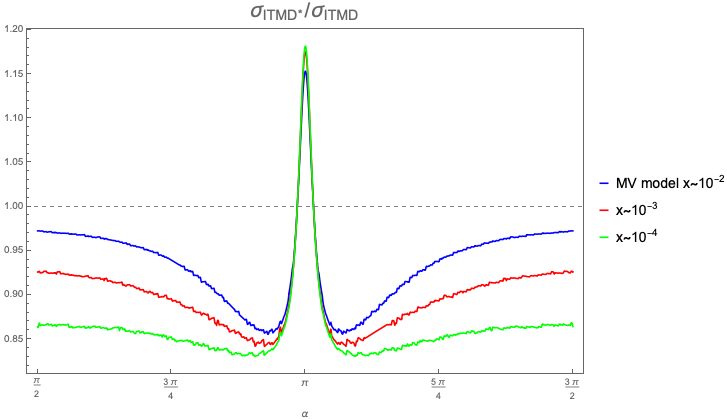} 
\caption{Ratio of the $pA\to Q\bar{Q}X$ cross sections calculated in the ITMD* and ITMD frameworks, as a function of the angle $\alpha$ between the transverse momenta of the heavy quarks. Gluon TMDs are numerically evaluated in the MV model, and evolved towards lower values of $x$ with the JIMWLK equations.}
 \label{fig:pA}
\end{figure}

The ratio between the ITMD* and ITMD cross sections for $pA\to Q\bar{Q}X$ is presented in fig.~\ref{fig:pA}:
\be\bea
\frac{\mathrm{d}\sigma_{gA}(\mathrm{ITMD}^*)}{\mathrm{d}\sigma_{gA}(\mathrm{ITMD})} &=\\&\hspace{-2cm}\frac{H_{1}^{ns}\big(\mathcal{F}_{gg}^{(1)}-\frac{1}{N_{c}^{2}}\mathcal{F}_{\mathrm{WW}}\big)+H_{\mathrm{ADP}}^{ns}\mathcal{F}_{\mathrm{ADP}}}{H_{1}^{ns}\big(\mathcal{F}_{gg}^{(1)}\!-\!\frac{1}{N_{c}^{2}}\mathcal{F}_{\mathrm{WW}}\big)
+H_{\mathrm{ADP}}^{ns}\mathcal{F}_{\mathrm{ADP}}+H_{1}^{h}\Big(\mathcal{H}_{gg}^{(1)}\!-\!\mathcal{F}_{gg}^{(1)}\!-\!\frac{1}{N_{c}^{2}}(\mathcal{H}_{\mathrm{WW}}\!-\!\mathcal{F}_{\mathrm{WW}})\Big)}\;.
\eea\ee
The deviation from the ITMD result is of the order of 15\%, and becomes more pronounced after evolution.

Clearly, the quality of the ITMD* approximation is process-dependent, which was to be expected given the fact that there is no small parameter which controls that approximation. The dramatic $\gamma^{*}_{\sL}A\to Q\bar{Q}X$ case looks to be an exception, due to the peculiar behavior of $H^{ns}_{\sL}$, but nevertheless even in those cases where the ITMD* part of ITMD is dominant, the ITMD*/ITMD ratio will be further deviating from unity with decreasing $x_{\scriptscriptstyle{A}}$, which is not satisfactory.

\section{On the diagrammatic approach to ITMD and its limitations}
In this section, we give an outline of the diagrammatic approach to ITMD and explain which problems arise when introducing a second hard scale. To do so, we focus on the simplest process that exhibits all the needed features: heavy-quark photoproduction.
\begin{figure}[t]
\centering
\includegraphics[width=7cm]{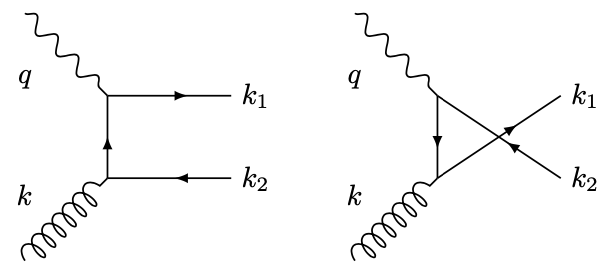} 
\caption{The two Feynman diagrams contributing to the heavy-quark pair photoproduction cross section.}
 \label{fig:Feynman}
\end{figure}
To set up the TMD calculation of this process, we work in a frame where the three-momenta of the right-moving photon and left-moving proton or nucleus lie on the $z$-axis. Using that $q^{2}=0$, $p_{{\scriptscriptstyle A}}^{2}=M_{{\scriptscriptstyle A}}^{2}\approx0$, and $s=(q+p_{\scriptscriptstyle{A}})^{2}\simeq2q\cdot p_{{\scriptscriptstyle A}}$, the light-cone plus- and minus directions are defined along $p_{{\scriptscriptstyle A}}^{\mu}$ and $2q^{\mu}/s$, respectively, such that a generic four-vector $p^{\mu}$ is decomposed as: 
\be\bea
p^{\mu}=(p^{+},p^{-},p_{{\scriptscriptstyle T}})=\frac{2p\cdot q}{s}p_{{\scriptscriptstyle A}}^{\mu}+p\cdot p_{{\scriptscriptstyle A}}\frac{2q^{\mu}}{s}+p_{{\scriptscriptstyle T}}^{\mu}\;.
\eea\ee
Using $k_{1}^{2}=k_{2}^{2}=m^{2}$, we then obtain for the other relevant momenta:
\be\bea
k^{\mu} & =xp_{{\scriptscriptstyle A}}^{\mu}+k_{{\scriptscriptstyle T}}^{\mu}\;,\\
k_{1}^{\mu} & =\frac{m^{2}+\mathbf{k}_{1}^{2}}{zs}p_{{\scriptscriptstyle A}}^{\mu}+zq_{{\scriptscriptstyle A}}^{\mu}+k_{1{\scriptscriptstyle T}}^{\mu}\;,\\
k_{2}^{\mu} & =\frac{m^{2}+\mathbf{k}_{2}^{2}}{\bar{z}s}p_{{\scriptscriptstyle A}}^{\mu}+\bar{z}q_{{\scriptscriptstyle A}}^{\mu}+k_{2{\scriptscriptstyle T}}^{\mu}\;,
\eea\ee
where $k_{{\scriptscriptstyle T}}^{2}=-\mathbf{k}^{2}$.

A standard calculation yields the following expression for the differential $\gamma A\to Q\bar{Q}X$ cross section:
\be\bea
\frac{\mathrm{d}\sigma}{\mathrm{d}z\mathrm{d}^{2}\mathbf{P}\mathrm{d}^{2}\mathbf{k}} & =\alpha\alpha_{s} e_{Q}^{2}\frac{z\bar{z}}{8(\mathbf{P}^{2}+m^{2}+z\bar{z}\mathbf{k}^{2})^{2}}\mathcal{M}_{\mu\nu}(z,\mathbf{k},\mathbf{P})\Gamma^{\mu\nu}(x,\mathbf{k})\;,\label{eq:photoCS}
\eea\ee
with the requirements that $\mathbf{k}=\mathbf{k}_{1}+\mathbf{k}_{2}$, $\bar{z}=1-z$, and $x=(\mathbf{P}^{2}+m^{2}+z\bar{z}\mathbf{k}^{2})/(z\bar{z}s)$. The perturbative hard part $\mathcal{M}_{\mu\nu}(z,\mathbf{k},\mathbf{P})$ encodes the amplitude squared of the two Feynman diagrams, with coupling constants, factors of $i$, and a factor $\frac{1}{4(N_{c}^{2}-1)}$ for the averaging over the spins and color of the incoming photon and gluon, taken out:
\be\bea
\mathcal{M}^{\mu\nu}(z,\mathbf{k},\mathbf{P})\equiv & -\mathrm{Tr}\Bigg[(\cancel{k}_{2}-m)\Big(\gamma_{\rho}\frac{\cancel{k}_{1}-\cancel{k}+m}{\hat{u}-m^{2}}\gamma^{\nu}+\gamma^{\nu}\frac{\cancel{k}_{1}-\cancel{q}+m}{\hat{t}-m^{2}}\gamma_{\rho}\Big)\\
 & \qquad\times(\cancel{k}_{1}+m)\Big(\gamma^{\mu}\frac{\cancel{k}_{1}-\cancel{k}+m}{\hat{u}-m^{2}}\gamma^{\rho}+\gamma^{\rho}\frac{\cancel{k}_{1}-\cancel{q}+m}{\hat{t}-m^{2}}\gamma^{\mu}\Big)\Bigg]\;.
\eea\ee
Note that the overall minus signs stems from the completeness relation of the photon polarization $\sum_{\lambda}\epsilon_{\lambda}^{\mu}(q)\epsilon_{\lambda}^{\nu*}(q)=-g^{\mu\nu}$. $\Gamma^{\mu\nu}(x,\mathbf{k})$ is the correlator of gluon fields:
\be\bea
\Gamma^{\mu\nu}(x,\mathbf{k}) & =\frac{2}{p_{{\scriptscriptstyle A}}^{-}}\int\frac{\mathrm{d}\xi^{+}\mathrm{d}^{2}\boldsymbol{\xi}}{(2\pi)^{3}}e^{i\xi^{+}k^{-}}e^{-i\boldsymbol{\xi}\mathbf{k}}\langle p_{{\scriptscriptstyle A}}|\mathrm{Tr}\,F^{-\mu}(0)\mathcal{U}(0,\xi^{+},\boldsymbol{\xi})F^{-\nu}(\xi^{+},\boldsymbol{\xi})|p_{{\scriptscriptstyle A}}\rangle\;,\label{eq:gluoncorrelator}
\eea\ee
which we parameterize, up to leading twist, in function of the unpolarized and linearly polarized gluon TMD \cite{Mulders:2000sh, Meissner:2007rx}:
\be\bea
\Gamma^{\mu\nu}(x,\mathbf{k}) & =-\frac{g_{{\scriptscriptstyle T}}^{\mu\nu}}{2}\mathcal{F}_{\mathrm{WW}}(x,\mathbf{k})+\Big(\frac{k_{{\scriptscriptstyle T}}^{\mu}k_{{\scriptscriptstyle T}}^{\mu}}{\mathbf{k}^{2}}+\frac{g_{{\scriptscriptstyle T}}^{\mu\nu}}{2}\Big)\mathcal{H}_{\mathrm{WW}}(x,\mathbf{k})\;,
\eea\ee
where 
\be\bea
g_{{\scriptscriptstyle T}}^{\mu\nu}\equiv g^{\mu\nu}-\frac{2}{s}\big(p_{{\scriptscriptstyle A}}^{\mu}q^{\nu}+q^{\mu}p_{{\scriptscriptstyle A}}^{\nu}\big)\;.
\eea\ee
The gauge link in eq.~\eqref{eq:gluoncorrelator} points towards future infinity due to the color flow of the process, and can be set equal to unity in a gauge choice $A^{-}=A^{+}=0$ and $A^{i}(\xi^{+}\to\infty)=0$, hence the gluon TMDs are from the WW kind.

From eq.~\eqref{eq:photoCS}, the TMD cross section is obtained by taking the back-to-back limit, i.e. contracting $\mathcal{M}$ with the tensors in $\Gamma$ and then taking the $\mathbf{k}\to0$ limit, such that the only $\mathbf{k}$-dependence left is in the TMDs, yielding:
\be\bea
\frac{\mathrm{d}\sigma}{\mathrm{d}z\mathrm{d}^{2}\mathbf{P}\mathrm{d}^{2}\mathbf{k}} & \overset{\mathbf{k}\to0}{=}\frac{\alpha\alpha_{s} e_{Q}^{2}}{(\mathbf{P}^{2}+m^{2})^{4}}\Big[\Big((\mathbf{P}^{4}+m^{4})(z^{2}+\bar{z}^{2})+2m^{2}\mathbf{P}^{2}\Big)\mathcal{F}_{\mathrm{WW}}(x,\mathbf{k})\\
 & \qquad+4z\bar{z}m^{2}\mathbf{P}^{2}\cos(2\varphi)\mathcal{H}_{\mathrm{WW}}(x,\mathbf{k})\Big]\;,
\eea\ee
The HEF result, on the other hand, is obtained by keeping nonzero $\mathbf{k}$ everywhere and taking the large-$\mathbf{k}^{2}$ limit of the gluon correlator:
\be\bea
\Gamma^{\mu\nu}(x,\mathbf{k}) & \to\frac{k_{{\scriptscriptstyle T}}^{\mu}k_{{\scriptscriptstyle T}}^{\mu}}{\pi\mathbf{k}^{2}}\mathcal{F}_{g/A}(x,\mathbf{k})\;,
\eea\ee
which gives:
\be\bea
\frac{\mathrm{d}\sigma}{\mathrm{d}z\mathrm{d}^{2}\mathbf{P}\mathrm{d}^{2}\mathbf{k}} & \overset{\mathbf{k}^{2}\gg Q_{s}^{2}}{=}\frac{\alpha\alpha_{s} e_{Q}^{2}}{\pi(\mathbf{k}_{1}^{2}+m^{2})(\mathbf{k}_{2}^{2}+m^{2})}\Big((z^{2}+\bar{z}^{2})+\frac{2m^{2}z\bar{z}\big(\mathbf{k}_{1}^{2}-\mathbf{k}_{1}^{2}\big)^{2}}{\mathbf{k}^{2}(\mathbf{k}_{1}^{2}+m^{2})(\mathbf{k}_{2}^{2}+m^{2})}\Big)\mathcal{F}_{g/A}(x,\mathbf{k})\;.
\eea\ee
As expected, the two above limits are in complete agreement with the TMD- and HEF-limits from the CGC calculation, which can be obtained by setting $Q\to0$ in the $\gamma^{*}A\to Q\bar{Q}X$ cross sections, eq.~\eqref{eq:CGCTMDgammaT} and eq.~\eqref{eq:HEFdis}.
Unfortunately, beyond these limits, the results from the diagrammatic approach become inconsistent. Indeed, we can cast the cross section eq.~\eqref{eq:photoCS} in a similar form as the ITMD result eq.~\eqref{eq:DISITMD} as follows:
\be\bea
\frac{\mathrm{d}\sigma}{\mathrm{d}z\mathrm{d}^{2}\mathbf{P}\mathrm{d}^{2}\mathbf{k}} & =\alpha\alpha_{s}e_{Q}^{2}\Big[\mathcal{M}^{ns}(z,\mathbf{k},\mathbf{P})\mathcal{F}_{\mathrm{WW}}(x,\mathbf{k})+\mathcal{M}^{h}(z,\mathbf{k},\mathbf{P})(\mathcal{H}_{\mathrm{WW}}(x,\mathbf{k})-\mathcal{F}_{\mathrm{WW}}(x,\mathbf{k}))\Big]\;,
\eea\ee
where:
\be\bea
\mathcal{M}^{ns}(z,\mathbf{k},\mathbf{P}) & =\frac{z\bar{z}}{8(\mathbf{P}^{2}+m^{2}+z\bar{z}\mathbf{k}^{2})^{2}}\mathcal{M}_{\mu\nu}(z,\mathbf{k},\mathbf{P})\frac{k_\sT^\mu k_\sT^\nu}{\mathbf{k}^2}\;,\\
\mathcal{M}^{h}(z,\mathbf{k},\mathbf{P}) & =\frac{z\bar{z}}{8(\mathbf{P}^{2}+m^{2}+z\bar{z}\mathbf{k}^{2})^{2}}\mathcal{M}_{\mu\nu}(z,\mathbf{k},\mathbf{P})\Big(\frac{k_\sT^\mu k_\sT^\nu}{\mathbf{k}^2}+\frac{g_\sT^{\mu\nu}}{2}\Big)\;.
\label{eq:photoprod}\eea\ee
One could now expect the above hard parts to be equal to the ones in eq.~\eqref{eq:DISITMDhard}. However, this turns out to be the case only for the non-sense projection:
\be\bea
\mathcal{M}^{ns}(z,\mathbf{k},\mathbf{P})&=H^{ns}_{\sT}(z,\mathbf{k},\mathbf{P})\;.
\eea\ee
The other projection obtained by the diagrammatic approach $\mathcal{M}^{h}$ does not match the ITMD result, except in the $\mathbf{k}\to 0$ limit: $\mathcal{M}^{h}(z,\mathbf{0},\mathbf{P})\!=\!H^{h}_{\sT}(z,\mathbf{0},\mathbf{P})$. This is because beyond this limit, projecting $\mathcal{M}_{\mu\nu}$ onto $g_\sT^{\mu\nu}$ is not gauge-invariant. So in general we have:
\be\bea
\mathcal{M}^{h}(z,\mathbf{k},\mathbf{P})&\neq H^{h}_{\sT}(z,\mathbf{k},\mathbf{P})\;.
\eea\ee
In particular, in the ITMD result obtained in this work, eq.~\eqref{eq:DISITMD}, the contribution from $H^{h}_{\sT}$ disappears in the $\epsilon_f^2\to 0$ limit. This is not the case, however, for the $\mathcal{M}^{h}$ projection calculated in the diagrammatic approach:
\be\bea
\mathrm{lim}_{m\to0}H^{h}_{\sT}(z,\mathbf{k},\mathbf{P})&=0 \;,\\
\mathrm{lim}_{m\to0}\mathcal{M}^{h}(z,\mathbf{k},\mathbf{P})&\neq0 \; ,
\eea\ee
signalling that, while the resumed ITMD hard factors $H^{ns}_{\sT}$ can be obtained following the simple diagrammatic calculation outlined in this section, this is not the case for their $H^{h}_{\sT}$ counterpart which pertains to the contribution of small-$x$ gluons that are not fully linearly polarized.

It was implicit that the ITMD hard factors discussed above are considered in the $Q^2\to 0 $ limit, since we are comparing them to \eqref{eq:photoprod}, valid for photoproduction. However the discussion remains valid in the $Q^2\neq 0$ case. Furthermore, it also applies in the case of heavy-quark production in $p+A$ collisions, channel by channel, i.e. 
$\mathcal{M}^{ns}_{(c)}(z,\mathbf{k},\mathbf{P})=H^{ns}_{(c)}(z,\mathbf{k},\mathbf{P})$ but $\mathcal{M}^{h}_{(c)}(z,\mathbf{k},\mathbf{P})\neq H^{h}_{(c)}(z,\mathbf{k},\mathbf{P})$.

\section{Conclusions}

In this paper we have explicitly calculated the ITMD factorization formulae for forward heavy-quark pair production in DIS and pA collisions. In the leptoproduction case
$(\gamma^* A\to Q \bar{Q} X$), the ITMD formula is \eqref{eq:DISITMD} with the hard factors \eqref{eq:DISITMDhard} obtained from \eqref{eq:decomphard} and \eqref{eq:HTns}-\eqref{eq:HTf} (transverse photon polarization) or \eqref{eq:HLns}-\eqref{eq:HLf} (longitudinal photon polarization). In the hadroproduction case $(g A\to Q \bar{Q}X$), the ITMD formula is \eqref{eq:sigmagA}, with the hard factors given by \eqref{eq:H1ns}, \eqref{eq:H1h} and \eqref{eq:Hadp}. All the gluon TMDs involved can be found in ref.~\cite{Marquet:2017xwy}, where they have been evaluated numerically (see e.g. figures 3 and 4 there). The ITMD formulae have been fully derived from CGC expressions (eq.(22) and (88) in \cite{Dominguez:2011wm}), they reduce to TMD expressions in the $Q_s,|\mathbf{k}| \ll |\mathbf{P}|$ limit where higher-twists are neglected, and to HEF expressions in the $Q_s \ll |\mathbf{k}|,|\mathbf{P}|$ limit where saturation effects are neglected.

We focused our attention to the $H^{h}_{(c)}(\mathcal{H}_{\mathrm{(c)}}\!-\!\mathcal{F}_{\mathrm{(c)}})$ terms.
Those terms disappear when either $H^{h}_{(c)}=0$, which happens when $m=0$ and $Q^2=0$ in which case we recover the original ITMD results of \cite{Kotko:2015ura}, or when $\mathcal{H}_{\mathrm{(c)}}\!=\!\mathcal{F}_{\mathrm{(c)}}$, which happens for TMDs of the dipole type, and in the HEF limit where all the gluon TMDs coincide.
However, in general the $H^{h}_{(c)}(\mathcal{H}_{\mathrm{(c)}}\!-\!\mathcal{F}_{\mathrm{(c)}})$ terms are not negligible, as shown in Figs.~\ref{fig:DIST},\ \ref{fig:DISL} and \ref{fig:pA}. We also showed that, while the HEF framework can be used to obtain the $H^{nc}_{(c)}$ off-shell hard factors by grouping diagrams into gauge invariant sub-sets \cite{Kotko:2015ura}, this diagrammatic approach is not, in its current form, able to reproduce $H^{h}_{(c)}$ ones.

Having in mind to extend the ITMD framework to processes with more than two particles in the final state, one would like to find a way to automatize the computation of the hard factors, either from the diagrammatic approach it if can be made to work (that automatization has been done for the calculation of the $\mathcal{F}_{\mathrm{(c)}}$ TMDs involved and of the $H^{ns}_{(c)}$ off-shell matrix-elements \cite{Bury:2018kvg,Bury:2020ndc}) or perhaps directly from the CGC by extending the approach of \cite{Boussarie:2020vzf}.

\section*{Acknowledgments}

We thank Renaud Boussarie and Farid Salazar for providing feedback on the manuscript. TA is supported by Grant No. 2018/31/D/ST2/00666 (SONA\-TA 14 - National Science Centre, Poland). This work has been performed in the framework of COST Action CA 15213 ``Theory of hot matter and relativistic heavy-ion collisions" (THOR), MSCA RISE 823947 ``Heavy ion collisions: collectivity and precision in saturation physics''  (HI\-EIC) and has received funding from the European Union's Horizon 2020 research and innovation programme under grant agreement STRONG2020 - No 824093.

\end{document}